\documentclass[a4paper]{spie}  %>>> use this instead for A4 paper

\usepackage[]{graphicx}
\usepackage{amssymb,amsmath,amsfonts}

\title{A report on the status of astrophotonics for interferometry and beyond}

\author{
Lucas~Labadie\supit{a}
\skiplinehalf
\supit{a} I.\,Physikalisches Institut, Universit\"at zu K\"oln, Z\"ulpicher Str. 77, 50937 K\"oln, Germany \\
}

\authorinfo{Further author information: (Send correspondence to L.Labadie)\\L.Labadie: E-mail: labadie@ph1.uni-koeln.de, Telephone:\,+49 221 470 3493
}
\begin{document} 
  \maketitle 

%%%%%%%%%%%%%%%%%%%%%%%%%%%%%%%%%%%%%%%%%%%%%%%%%%%%%%%%%%%%% 
\begin{abstract}
Long-baseline interferometry and high-resolution spectroscopy are two examples of areas that have benefited from astrophotonics devices, but the application range is expanding to other subareas and other wavelength ranges. The VLTI has been one of the pioneering astronomical infrastructure to exploit the potential of astrophotonics instrumentation for high-angular resolution interferometric observations, whereas new opportunities will arise in the context of the future ELTs. 
In this contribution, I review the current state of the art regarding the interplay between photonic-based solutions and astronomical instrumentation and highlight the growth of the field, as well as its recognition in recent strategy surveys such as the Decadal. I will explain the benefits of different technological platforms making use of photolithography or laser-writing techniques. I will review the most recent results in the field covering simulations, laboratory characterization and on-sky prototyping. Astrophotonics may have a unique role to play in the forthcoming era of new ground-based astronomical facilities, and possibly in the field of space science.
\end{abstract}

%>>>> Include a list of keywords after the abstract 


%%%%%%%%%%%%%%%%%%%%%%%%%%%%%%%%%%%%%%%%%%%%%%%%%%%%%%%%%%%%%
\section{Introduction}

In the era of modern astronomy and astrophysics, imaging and spectroscopy are two major pillars
to observe and study our Universe. 
To date, there is little debate on the critical importance of
technical and technological advances to enable new, unprecedented discoveries that can fundamentally
deepen our understanding of the Universe. This technological endeavour, without which observational
capabilities would stagnate, addresses important challenges relevant to instruments and facilities
such as, among many other examples, building large telescope apertures and interferometers, imagers delivering high-angular resolution and high-contrast capabilities, high-precision and
high-stability spectroscopic units and state-of-the-art large area low-noise detectors. 
The steady growth of multi-messenger astronomy requires collecting data across the whole electromagnetic spectrum, which would remain impossible without a continuous growth of astronomical instrumentation. \\
The last decade has seen tremendous progress in several areas of astrophysics, which has been supported by steadily improving instruments at, e.g., the VLT/VLTI or ALMA, whereas the James Webb Space Telescope launched in December 2021 will deliver transformational science in the near future. 
In this context, the field of astrophotonics has steadily grown at the interface between photonics and
astronomical instrumentation. 
Initially expanding from the established field of optical telecommunication technologies tailored to astronomical applications, the astronomy-photonics synergy has led to the emergence of novel breakthrough concepts, which maturity ranges from laboratory demonstrators to operational community-wide astronomical instruments.

\noindent Since the SPIE meeting on astronomical instrumentation in 2016 in Edinburgh, the organizers of the conference on optical and infrared interferometry have regularly included in the program a specific session on astrophotonics technologies to be introduced by a review of the field \cite{Labadie2016,Norris2018,Cvetojevic2020}\,. This illustrates the increasing scientific and technical  relevance of astronomical photonics for interferometry, as seen in the success of the VLTI with GRAVITY or with new key instrumentation at CHARA, all using guided optics building blocks. Astrophotonics has already more than two decades of history through the fruitful collaboration of various communities that were motivated by the development of new astronomical instruments operating primarily in the visible and infrared range, and for which unique optical functionalities could be enabled thanks to photonic-based designs not accessible with a more classical bulk optics approach.

\noindent From an historical perspective, astrophotonics represents a tiny fraction of the multi-century long expansion phase of instrumentation in astronomy, primarily because of the recent nature of the elementary photonic components that are at the heart of the discipline. But revolutionary discoveries in astronomy and astrophysics have always been the fruit of patient technological and engineering innovation. And indeed, for many of us in the community, it is still fascinating to remember that the legacy of Edwin Hubble and its groundbreaking transformation of our understanding of the size and nature of the Universe is intimately linked to the tireless engineering effort of a group a persons who managed to build, assemble and maintain the Hooker telescope and its 100-inch primary mirror on Mount Wilson, the world's largest optical telescope in the first half of the twentieth century. The power of which was further expanded by Albert Michelson using his boom innovation with which he measured Betelgeuse's diameter. Well, astrophotonics has not had this influence yet, but its role in transforming modern optics for astronomy cannot be underestimated. 

\noindent In Section 2, I present the context of the astrophotonics activities and emphasizes the growth of the community. In Section 3, I detail the so-called ``Astrophotonic flow'' and uses it to present recent achievements in the field, focusing primarily on the topic of the conference, namely interferometry. In Section 4, I discuss few bright perspectives for our community.

%----------------
\begin{figure}[t]
\centering
\includegraphics[width=\textwidth]{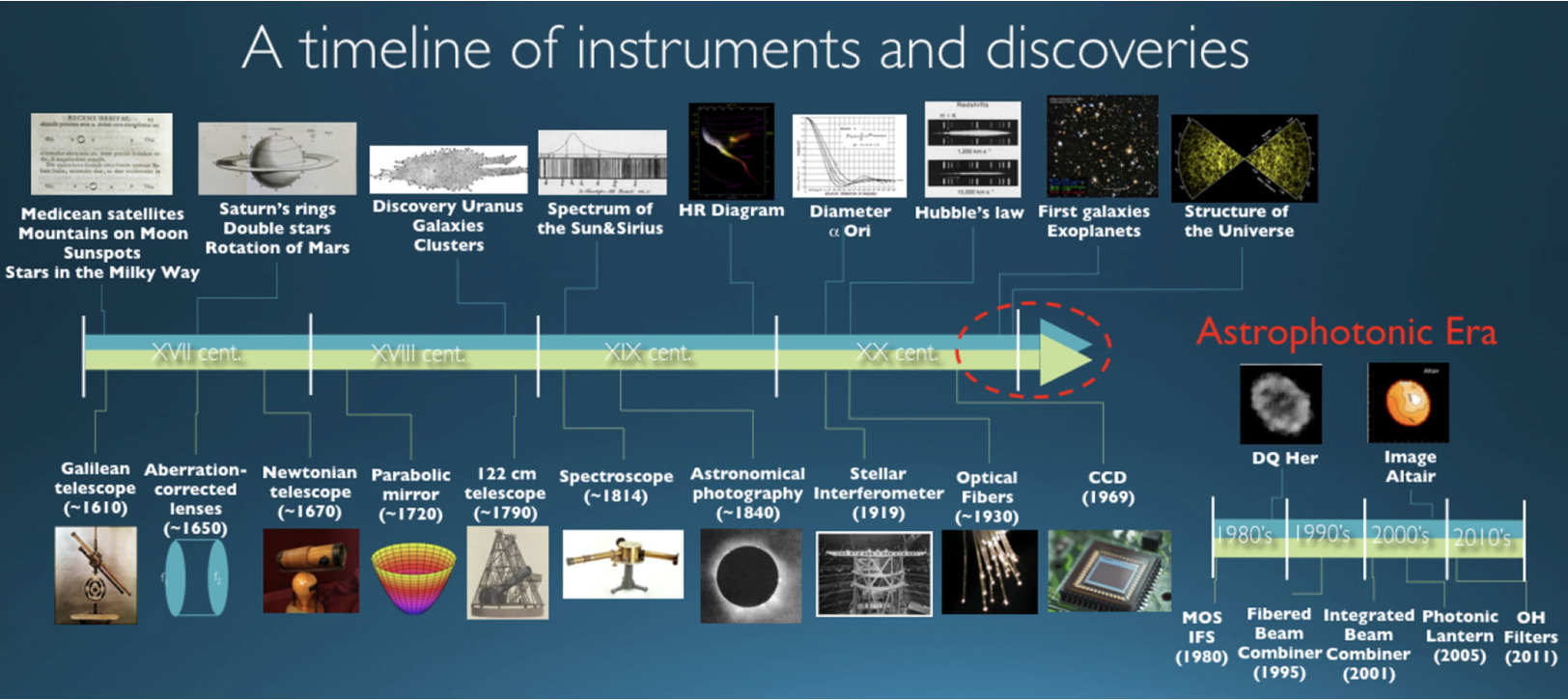}
\caption{An historical perspective to the synergy between instrumentation and observational astronomy, or how new technologies and fundamental discoveries go hand in hand. From Minardi, Harris \& Labadie, 2021 \cite{Minardi2021}\,.}\label{fig1}
\vspace{0.0cm}
\end{figure}
%----------------

\section{A growing and active community}

\begin{figure}[t]
\centering
\includegraphics[width=\textwidth]{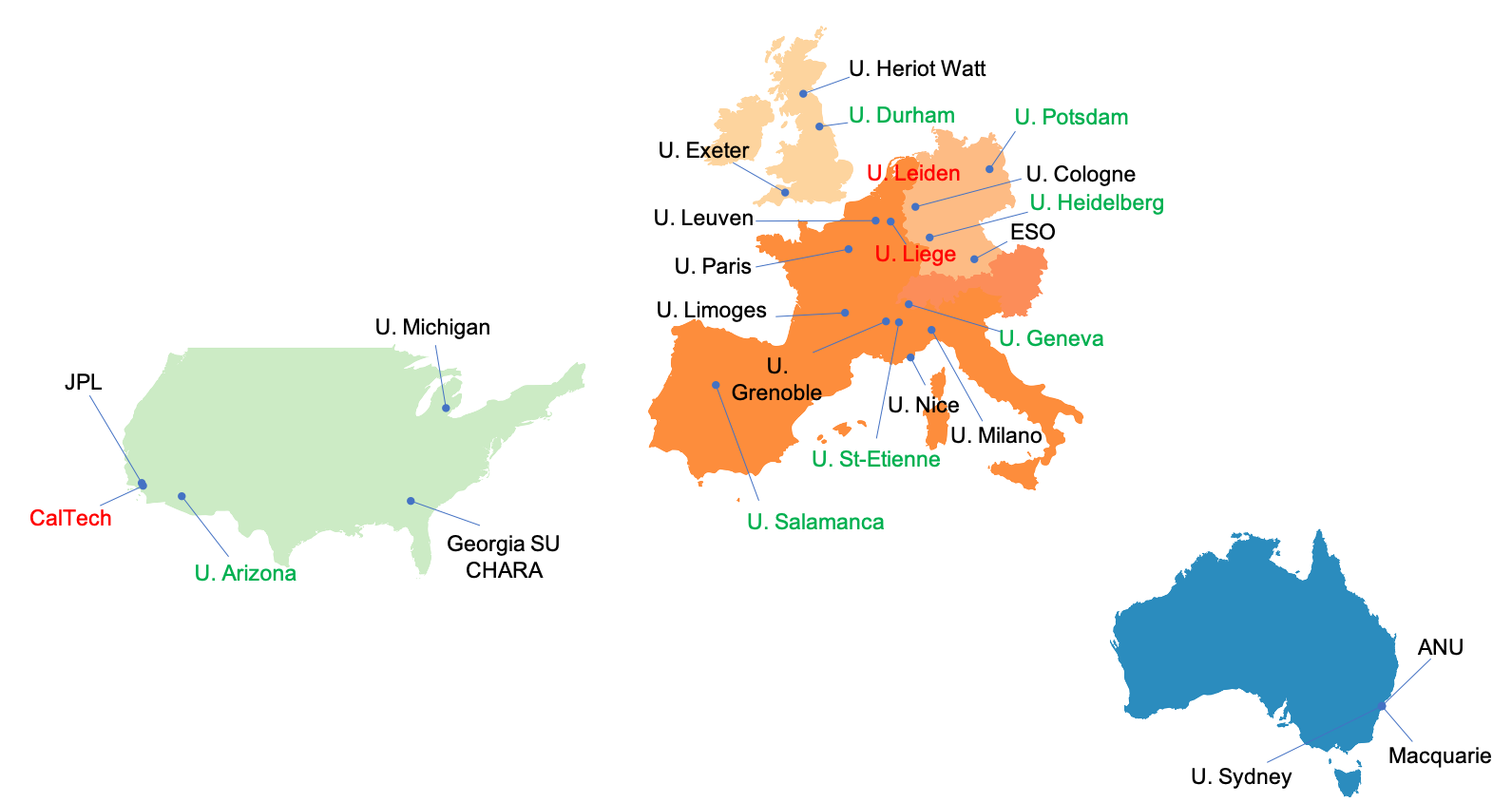}
\caption{Tentative visualization of the geographical repartition of research groups active in astrophotonics. 
The graph does not claim full exhaustivity. 
The color code attempts to coarsely identify the observing technique of main interest of the group, with black assigned to interferometry, green to high-resolution spectroscopy, and red to high-contrast imaging. I emphasize the limits of this approach since many institutions listed here may target several techniques. As an example, the University of Sydney has groups with recognized contributions in astrophotonics applications for both high-resolution spectroscopy and interferometry.}\label{fig2}
\vspace{0.0cm}
\end{figure}

While this review concentrates primarily on interferometry as the main topic of the conference, it is noticeable that the applications of astrophotonics go well beyond this community, triggering the interest of scientists that need to overcome major limitations of existing observational capabilities. In brief, the techniques of long-baseline interferometry, high-contrast imaging and high-resolution spectroscopy represent the three major areas relevant to the field of astrophotonics. At the time of writing, the majority of the groups active in the field or showing interest in its exploitation are located in Europe, with the largest concentration in France, Germany and the United Kingdom. The European groups are active in all areas of astrophotonics applications to the aforementioned observing techniques. A smaller albeit very active and diversified community is located in the area of Sydney, in Australia. Finally, groups with a marked interest in high-contrast imaging techniques and high-resolution spectroscopy are found in the United States, alongside with groups historically involved in the implementation of photonics for interferometry. Most groups identified in Fig.~\ref{fig2} have established close collaborations with highly complementary competencies in the areas of fabrication, characterization and integration of astrophotonics components. All share the common objective to exploit astrophotonics solutions either to increase the stability and reduce the overall complexity of an instrument compared to its bulk optics counterpart, or to enable optical functions that can only be reliably enabled with the help of photonics. In this last category, I consider devices such as the fibre Bragg gratings for the efficient suppression of the narrow and bright hydroxyl sky emission lines in the near-infrared \cite{Bland-Hawthorn2011}\,, or the laser-written waveguide networks used to reconfigure a pupil or an image plane, as implemented for instance in the Dragonfly instrument \cite{Jovanovic2012}\,.

%----------------
%----------------

\noindent Astrophotonics is to be considered as an active research field in which fundamental questions related to the phase control of a wavefront \cite{Ellis2021}\,, the exploitation and optimization of manufacturing processes \cite{Gatkine2021}\,, or the quantum manipulation of single photons \cite{Bland-Hawthorn2021} are at the heart of a vivid academic activity. Recently, a special feature focusing on astrophotonics has been jointly published by two journals of the Optical Society of America with about thirty papers broadly addressing the progress and challenges of photonics in astronomy \cite{Dinkelaker2021a,Dinkelaker2021b}\,. Hence, as an emerging academic field, astrophotonics goes beyond the simple exploitation of the existing photonics market, but rather invents new concepts and proposes new ideas serving astronomical needs and rarely emerging from the conventional photonic community. The relevance of astrophotonics for the astronomical community as large and its recognition as an academic field is probably best illustrated with the first recently invited review of the field in the Astronomy \& Astrophysics Review \cite{Minardi2021}\,.

\begin{figure}[t]
\centering
\includegraphics[width=\textwidth]{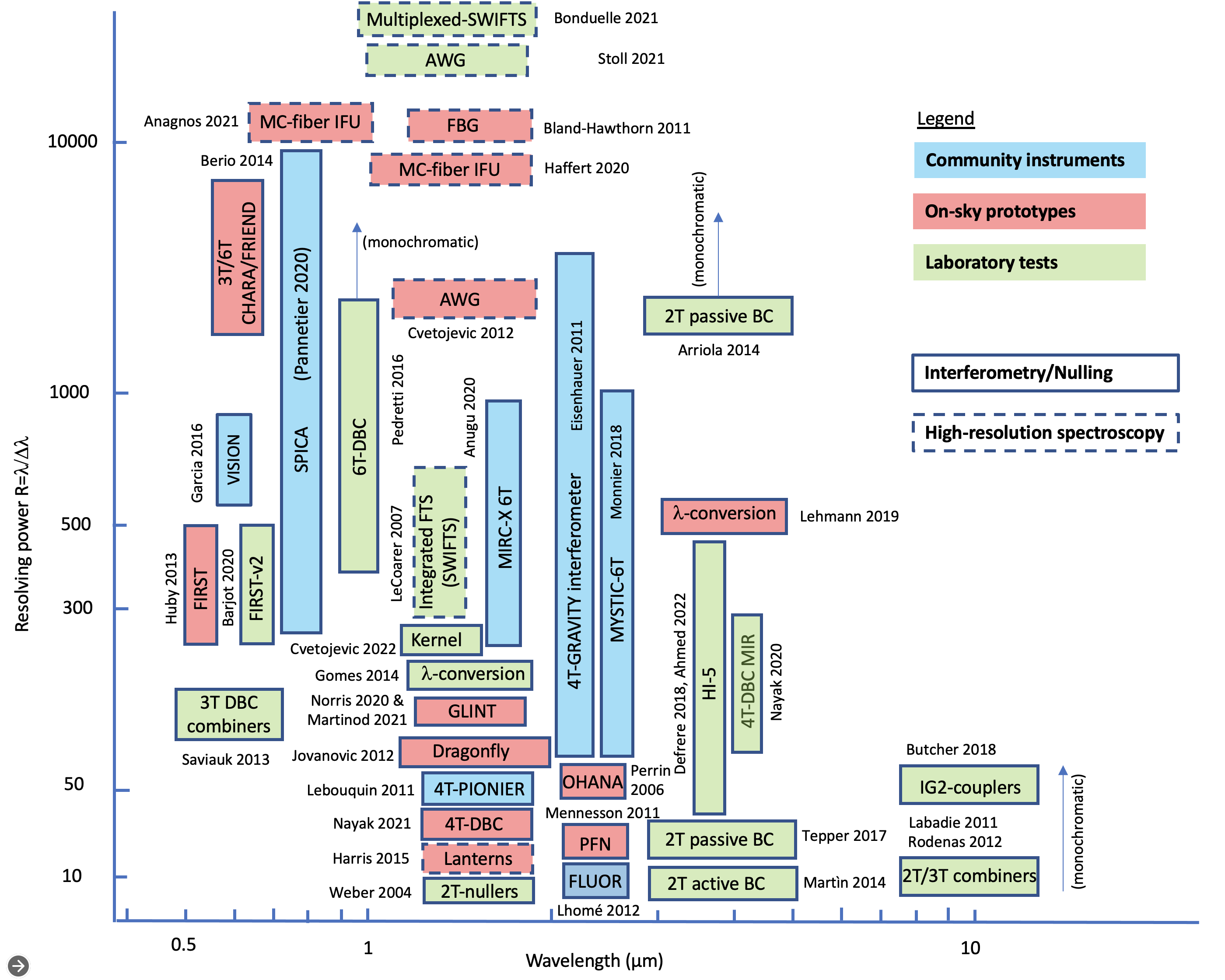}
\caption{
Yield of the astrophotonics venue in 2022. The technique of high-contrast imaging involving the definition and development of phase masks is not shown here.}\label{fig3}
\vspace{0.0cm}
\end{figure}

\noindent In an invited paper published in 2016, I represented the yield of the astrophotonics venue for different observational techniques such as interferometry and high-resolution spectroscopy in the form of a graph locating different projects and initiatives in a ``spectral-resolution versus wavelength-coverage'' plot (see Fig.\,7 in Labadie et al. 2016 \cite{Labadie2016})\,. In particular, I differentiated the level of maturity by grouping them in the categories of lab demonstrators, on-sky experiments or prototypes and community instruments. In Fig.~\ref{fig3}, I extended this view taking into account a six-year span. The comparison clearly illustrates how much significant progress has been achieved. Following the immense success of the GRAVITY instrument at the VLTI, other community instruments are now offered mainly as single-mode interferometers, as for instance the complementary pair MIRC-X/MYSTIC sharing the CHARA platform \cite{Monnier2018,Anugu2020}\,, or the commissioned visible SPICA interferometer \cite{Pannetier2020}\,
%({\bf Mourard 2022, this conference proceeding})
, all combining six telescopes. In the upper part of the plot, one can observe clear progress in the area of photonic-based functionalities for high-resolution spectroscopy. While still at the stage of lab experiments, the silica-on-silicon (SiO$\rm 2$/Si) lithographic platform has been able to produced arrayed waveguide gratings (AWGs) demonstrating a spectral resolution of almost 30,000 in the H\,band over a bandwidth larger than 100\,nm \cite{Stoll2021}\,, which is to my knowledge about the highest spectral resolution demonstrated with AWGs. Similarly, in the field of static Fourier Transform spectroscopy (SWIFTS), Bonduelle et al. \cite{Bonduelle2021} has demonstrated experimentally the successful multiplexing of the SWIFTS concept with no movable parts to reconstruct a near-infrared spectrum with a resolution larger than $R$\,=\,30,000. Further slightly deviating from the applications of astrophotonics to interferometry, it is of high interest to report the on-sky demonstration of the concept of MCF-IFU (i.e. Multi-Core Fiber Integral Field Unit) at visible and near-infrared wavelengths by Anagnos et al. \cite{Anagnos2021} and Haffert et al. \cite{Haffert2020}\,, which emerges as a promising extension of the well established principle of fiber fed spectrograph. I will briefly come back to this idea in Sect.~\ref{Remapp-printing}.
%----------------
%----------------
\\
%----------------
%----------------
It is striking to remark that the research in astrophotonics remains strongly dominated by devices operating in the near-infrared range, and more particularly around 1550\,nm, which reminds us the strong heritage of the telecommunication and semi-conductor fields despite active research programs to steer photonics towards more specific needs for astronomy. In fact, a significant effort has been undertaken to open the mid-infrared window to long-baseline interferometry and the results obtained by Tepper et al. \cite{Tepper2017a,Tepper2017b} and Gretzinger et al. \cite{Gretzinger2019} are now converging towards the construction of Hi-5 instrument \cite{Defrere2018}\,, the first integrated-optics based mid-infrared instrument at the VLTI.

\section{The astrophotonics flow}

\begin{figure}[t]
\centering
\includegraphics[width=\textwidth]{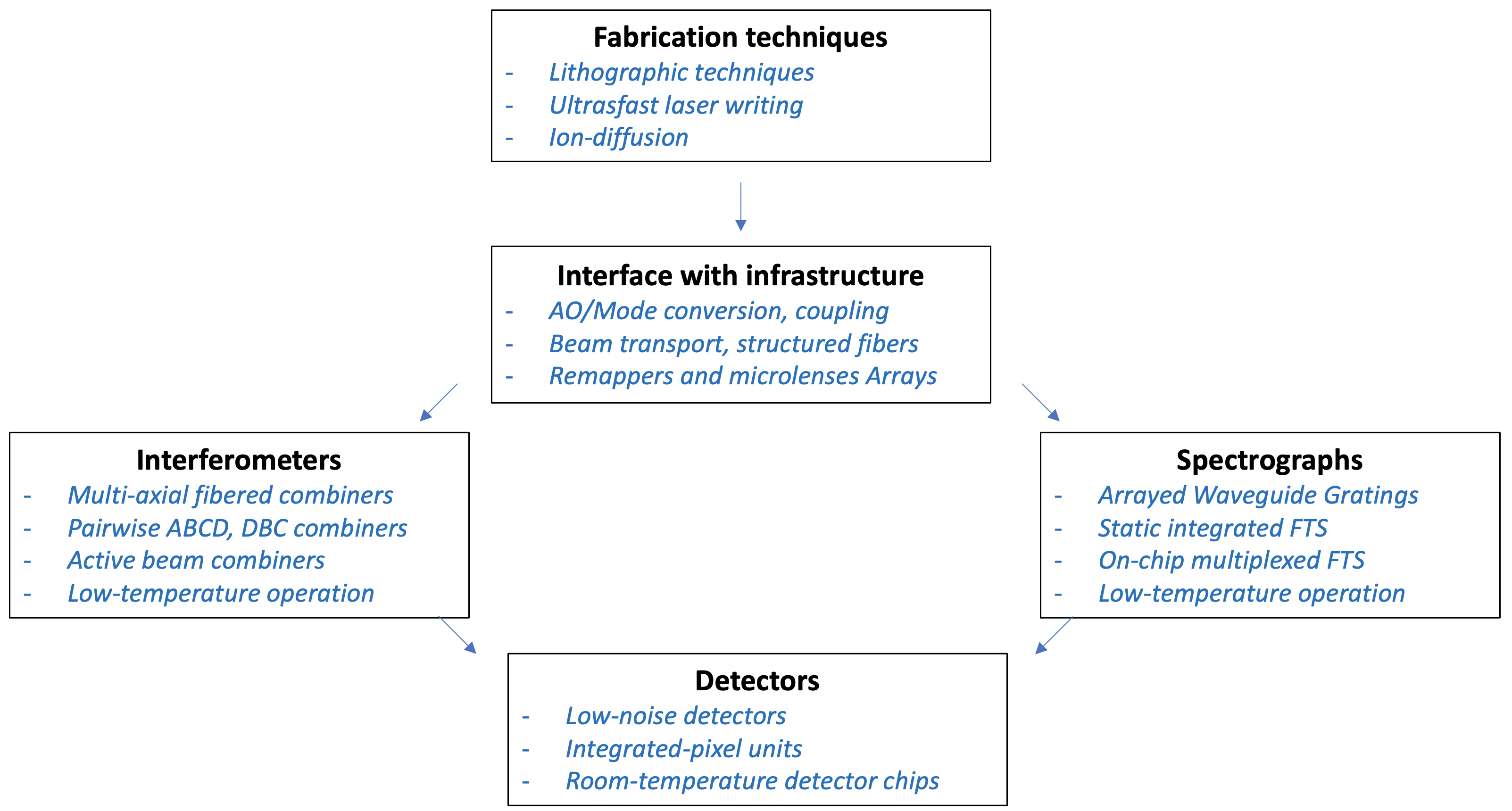}
\caption{
A possible development flow for a photonic-based astronomical instrument.}\label{fig4}
\vspace{0.0cm}
\end{figure}

The development of an astronomical instrument is always primarily motivated by the scientific objectives to be achieved, which are then broken down into technical requirements. In case the photonic option is considered more competitive than the bulk optics one -- in particular with respect to recognized advantages such as compactness, instrumental stability, cost and potential maturity -- a development flow can be established as illustrated in Fig.~\ref{fig4}. Note that for the purpose of this article, only the cases of interferometry and spectroscopy are considered. The problematic of the detectors together with the techniques of signal recording and processing is not discussed here. 

%----------------
%----------------
%----------------
%----------------

\subsection{Manufacturing aspects}

As a first consideration, the availability and maturity of the fabrication platform needs to be assessed. The choice is motivated by the high-level requirements of the instrument such as the spectral range of operation, the level of complexity of the optical functions to be implemented, and the expected throughput or insertion losses. The three main fabrication platforms that can be currently considered are the lithography platform (e.g., photolithography or E-beam lithography), the ultrafast laser writing (ULI) platform, and the ion-diffusion in glasses. It is outside the scope of this review to describe in details the underlying processes and methods for each platform and I refer the reader to the large existing literature on the topic or to the corresponding references in Labadie et al. \cite{Labadie2016} as a starting point. I only point out that I do not consider here the potential of these platforms for instrumentation in the UV spectral range or for wavelengths longer than $\sim$10\,$\mu$m as little has been done with astrophotonics in these domains. Few important points can be nonetheless highlighted.
\begin{itemize}
	\item For all the three platforms, the propagation losses can be grossly estimated to less than a dB/cm, in particular in the near-infrared range around 1550\,nm where silica-based photonics exhibits losses smaller than 0.1 dB/cm. In general, the propagation losses increase significantly when using more exotic materials for the mid-infrared range (see Butcher at al. \cite{Butcher2018} at 7.85\,$\mu$m).  
	\item The level of field confinement is directly dependent on the achievable index contrast, which in return determines the possible level of compactness of the chip due to more or less important bending losses. Currently, the silica-based platform (e.g. SiO$_{\rm 2}$/Si) offers the best option for high index contrast with achievable $\Delta$n as high as 0.1.
	\item Generating sub-micron photonic structures is at reach with, for instance, e-beam lithography. However, it was recently demonstrated by that the versatile laser-writing platform is also capable of generating $\sim$100\,nm-scale isotropic nanovoids that could be exploited for the development of low-resolution dispersing elements \cite{Lei2021}\,. This might be a future interesting perspective for astrophotonics.
\end{itemize}
	  
\subsection{Interface to the infrastructure}\label{interface}

A critical part of the flow concerns the interface with the infrastructure, whether a telescope or an interferometric array. How the light is collected at the focus of the telescope(s) and how it will be injected into a photonic chip stage? 

\subsubsection{Mode control}
For single apertures, most recent photonic instruments have emphasized the importance of operating in conjunction with an adaptive- or even extreme adaptive optics system \cite{Jovanovic2016}\,. Indeed, the precise phase control of the incoming wavefront through modal filtering or the circumventing of the spectrograph-telescope size relation requires to operate in the single-mode regime, which implies coupling a diffraction-limited spot in the input waveguide. In this sense, the increasing availability of adaptive optics on large telescopes helps significantly the usage of single-mode photonic devices. \\
In case of a modest correction from the AO stage, mode conversion using a photonic lantern has been proposed\cite{LeonSaval2005,Thomson2011} and can be implemented in the few-mode regime or in the seeing-limited regime for which the number of modes is approximated\cite{Harris2015}
%\begin{eqnarray}
by the relation 
M$\sim$($\pi\alpha$D/4$\lambda$)$^2$, where $M$ is the number of modes, $\alpha$ the angular size of the PSF and $D$ the telescope diameter, all in IS units. With the mode converter, the idea is to attack the photonic device always in the single-mode regime. While the mode converting photonic lantern has found applications for high-resolution spectroscopy\cite{Bland-Hawthorn2011,Harris2015}\,, its implementation for long-baseline interferometry is less straightforward. The main reason is that a lantern interfaced with a few-mode or seeing-limited point-spread function delivers single modes with a random time-variable phase relationship between them, which is equivalent to multimode interferometry that essentially scrambles the measured contrast. This is the reason why all optical/infrared interferometers have equipped the array telescopes with adaptive optics, the last example being the NAOMI adaptive optics system on the VLTI auxiliary telescopes\cite{Woillez2019} replacing the single tip-tilt system. Note however that recent progress on a all-photonic wavefront sensor by Norris et al.\cite{Norris2020} may change this perspective.

\subsubsection{Beam transport}

\begin{figure}[t]
\centering
\includegraphics[width=0.65\textwidth]{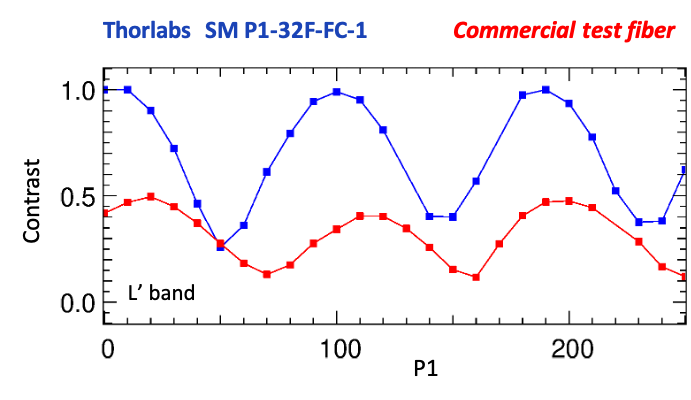}
\caption{
Plot giving in y the polarization contrast of an output field versus the angular direction of a linearly polarized input field, in degrees. The entrance polarizer placed before the fiber forces the angle of the input linear polarizer, while the exit analyzer measures the polarization state. A contrast of one indicates a linear polarization, a contrast of zero indicates a circular polarization. The measurement is done at 3.8\,$\mu$m.
 }\label{fig5}
\vspace{0.0cm}
\end{figure}

In long-baseline interferometry, optical fibers have long played an important role in interfacing the individual apertures with the beam combiner. It is generally done only at the level of the instrument over few meters, like at the VLTI or at CHARA, where an upstream classical optical train forms the optical relay and the delay lines for the beam transport in the interferometry combination lab. In this configuration, near-infrared fibers are now well developed with exceptionally low losses of $\sim$0.2\,dB/km at 1550\,nm. High-quality Nufern\,1950 fibers cover the K-band as well with polarization-maintaining capabilities. Commercial polarizing (PZ) fibers -- only transmitting one linear polarization whatever the input polarization state -- are also available in the near-infrared, as well as endlessly single mode fiber and large-mode area fibers. In other words, optical fibers for the H\,band -- and to a large extent for the K\,band as well -- do not represent any particularly strong challenge for astrophotonics.\\
Fiber transport over much larger distances, hence replacing the bulk optics optical train, has been of the interest of interferometrists for a long time, originally starting with the OHANA experiment\cite{Perrin2006b} in the K band and combining the two Keck telescopes. To my knowledge, this is to date the only example of hectometric fibered links for direct detection interferometry. Another noticeable example regards the recent implementation of hectometric fiber links between the CHARA telescopes to support accurate stabilization of the optical path difference down to 4\,nm rms using stretchers of polarization-maintaining fibers operating at 1550\,nm\cite{Lehmann2019}.\\
% Building in part on this heritage, a recent project named STELLIM revises this approach by using optical fibers in the visible for beam transportation from about ten small telescopes er of sub-apertures 
At wavelengths longer than 2.2\,$\mu$m, the situation appears less mature. Mid-infrared fluoride-based solid-core fibers are now commercially available on the market, but with poorer performance than the silica-based counterparts. To date, the basic entry-level Thorlabs fibers exhibit ``only'' 0.5dB/m ($\sim$90\% throughput over 1\,m), which remains a relatively modest value, while the polarization properties remain generally insufficiently constrained. In Fig.~\ref{fig5}, I show a comparison of the polarization behavior for an unspecified Thorlabs fiber and a second commercial fiber announced to be polarization-maintaining, with clearly different results indicating the low level of maturity in treating the polarization properties of these mid-infrared fibers. The Thorlabs fiber (blue curve) shows clear angular directions separated by 90$^{\circ}$ for which the input linear polarization remains unchanged, hence identifying the fast and slow axis reported usually for a polarization-maintaining (PM) fiber. On the contrary, the measurement on the anticipated PM test fiber indicates a scrambling of any input linear polarization. Of course, these measurements do not set any sort of quantitative classification between manufacturers but simply suggest, based on systematic tests, the difficulty to have the polarization state of the field through the fiber transmission correctly specified.
%----------------
%----------------
\\
%----------------
%----------------
In the longer wavelength range corresponding to 10\,$\mu$m, the state of the art remains unfortunately even more primitive. Hollow-core, large area, quasi single-mode and multimode are offered commercially with propagation losses on the order of 0.5\,dB/m as well, whereas chaclogenide fibers covering the 1-6\,$\mu$m range and polycristalline fibers covering the 4-17\,$\mu$m are produced, though to my knowledge none of these fibers are reported to having been used for astronomical applications, yet.

\subsubsection{Remapping devices and microlenses}\label{Remapp-printing}

\begin{figure}[t]
\centering
\includegraphics[width=\textwidth]{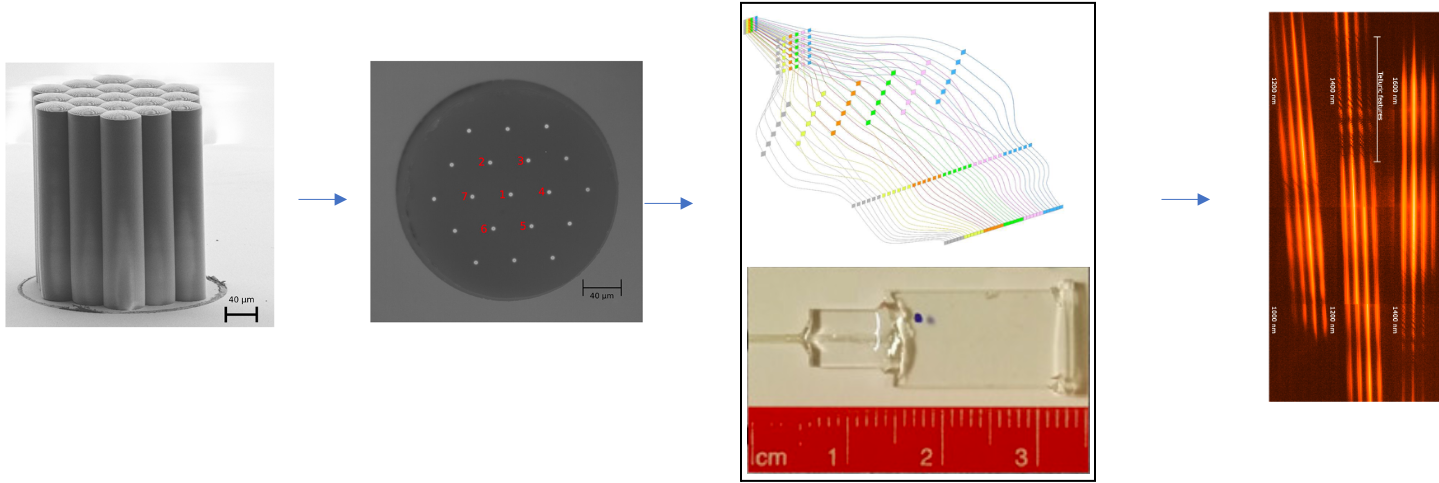}
\caption{Illustrating flow of a Multicore-Fiber Integral Field Unit (MCF-IFU). This image is a collage using data from Anagnos et al.\cite{Anagnos2021}, Harris et al.\cite{Harris2015} and Haffert et al.\cite{Haffert2020}\,. From left to right one sees the 3D-printed lenslets requiring Gaussian beam optics treatment, the input of the multicore fiber, the remapping lantern to form the pseudo-slit feeding a spectrograph covering the J and H bands in this case.}\label{fig6}
\vspace{0.0cm}
\end{figure}

The last interface unit between the telescope and the photonic chip that I wish to address concerns the reformmatting and/or sampling of the pupil or image plane. In pupil-remapping based aperture masking experiments, photonic technologies have demonstrated their potential in breaking the baseline redundancy of a full telescope pupil, while preserving the advantage of the aperture masking technique capable of detecting a faint companion close or even below the $\lambda$/D resolution limit\cite{Biller2012}\,. The FIRST instrument\cite{Huby2012,Huby2013} represents a classical example of fiber-linked remapper, but noticeable progress has been obtained with ultrafast laser writing of three-dimensional pupil remappers in a single and compact glass substrate, which typically guarantees high mechanical and thermal stability in comparison to a fiber network. Pioneered in the DRAGONFLY instrument\cite{Jovanovic2012}\,, it is also used in the GLINT nulling instrument\cite{Norris2020b} as well as in the four-telescope Discrete Beam Combiner experiment\cite{Nayak2021} (see Sect.~\ref{interfero}). In this last three cases, the exploitation of the ULI technique has been key in obtaining these results.\\
While the pupil plane can be sampled using photonic devices, the same can occur with the image plane of a telescope. I already mentioned here above the implementation of a photonic lantern in the image plane to serve as all-photonic wavefront sensor\cite{Norris2020}\,, which shares a similar conceptual approach -- although different in its implementation -- to the micro-lens tip-tilt sensor of Hottinger et al.\cite{Hottinger2021}\,. In the area of telescope interfaces with image plane sampling/remapping, a promising result was recently published where a small portion of the extreme-AO corrected field-of-view with SCExAO on the 8-m Subaru telescope was sampled and remapped into a pseudo-slit feeding a high-resolution spectrograph. The all-photonic optics train -- except for the dispersing element -- was based on a multi-core fiber (MCF) on top of which a microlens array was 3D-printed to sample the image plane. At the extreme end of the MCF, the cores were rearranged from a 2D to a 1D linear arrangment using a remapping lantern. This prototype was optimized for the typical astrophotonic H-band\cite{Anagnos2021}\,, whereas a visible version was assembled for the 4.2-m William Herschel Telescope\cite{Haffert2020}\,. While the concept of image plane sampling using fiber bundles or bulk optics lenslet arrays is not new with these works, it is interesting to demonstrate the feasibility of a one-block interface with a filling factor of the image plane of about 100\%. The 3D-printed lenslet approach remains however still limited in its physical extension, which in returns limits the total size of the sampled field-of-view. 
I illustrate in Fig.~\ref{fig6} the original interface flow between the telescope and the spectrograph. A possible extension of the 3D-printed lenslets towards the mid-infrared might be accessible in the future, albeit depending on the progress that can be achieved in the area of polymer resins transparency.

%----------------
%----------------
%----------------
%----------------

\subsection{Astrophotonics in Interferometry}\label{interfero}

Following the flow of Fig.~\ref{fig4}, once the interface with the telescope or the interferometric array is ensured (see Sect.~\ref{interface}), the next step is to address the astrophotonics design of the beam combination stage: which combination scheme is more appropriate to the high-level science requirements of the instrument? What is the targeted sensitivity ultimately justifying the use of photonic elements? How the imaging capabilities (e.g., the number of apertures) connects with the technical complexity of the beam combiner? Is a stage of rapid phase control required? What about the need for low-temperature operations? In this context, I address in the following paragraphs the progress in astrophotonics for the field of interferometry.

\subsubsection{Fiber-based beam combiners}\label{fiberbased}

Using single-mode fibers to implement a multi-axial interferometric beam combination is the lowest level of complexity involving astrophotonics components. In several cases, non-redundant multi-axial fiber-based interferometers are actually referred to as bulk-optics concepts. To date, all interferometric instruments that have adopted this scheme are located at the CHARA interferometric facility, namely MIRC-X, MYSTIC and VEGA. It is also remarkable that this is so far the only scheme adopted for the interferometric combination of six telescopes. 
% Do you need single-mode? 

\subsubsection{ABCD Integrated-optics (IO) beam combiners}\label{io-beamcombiners}

\begin{figure}[t]
\centering
\includegraphics[width=0.55\textwidth]{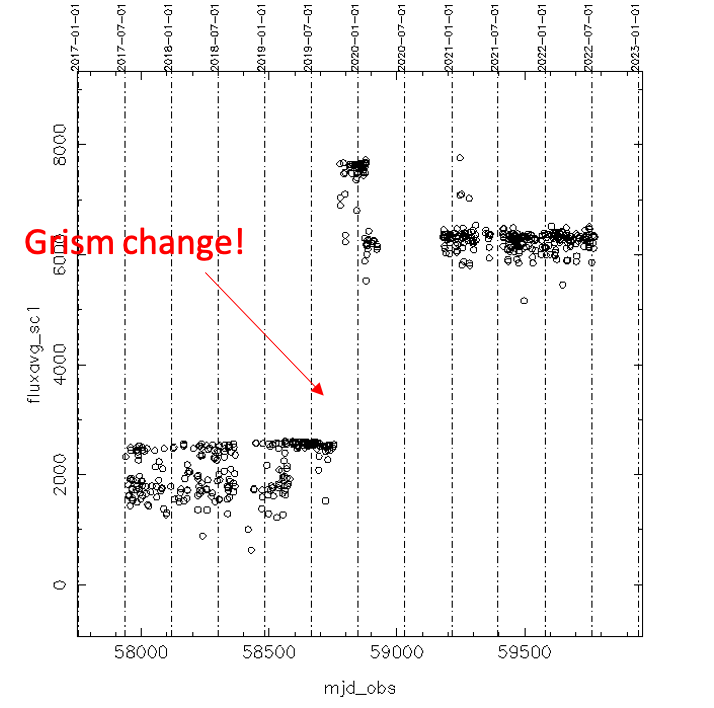}
\caption{Overall throughput of the GRAVITY instrument and its integrated beam combiner after five years of operation at low temperature. See text for details.
}\label{fig7}
\end{figure}

The GRAVITY experience -- and success -- is based on a silica-on-silicon platform beam combiner with four inputs and 24 outputs to implement the static ABCD combination scheme to encode the interferometric quantities and the telescopes fluxes. This chip, well-known to the community, and its principle are described in details elsewhere\cite{Benisty2009,Perraut2018}\,. Here, I will only concentrate on the fact that the GRAVITY beam combiner is the only photonic chip that has operated in low-temperature conditions ($\sim$\,-80$^{\circ}$C) over the last five years. It can be therefore of high interest to verify the survival conditions of this chip, and in particular of its glued connections to the fibers. Since it is not possible to remove the IO beam combiner from GRAVITY for test, and since to my knowledge no ``ground model'' exists in the same operational conditions of the ``flight model'', we can analyze the evolution of the throughput of the instrument to at least exclude that the beam combiner is the worst offender in terms of throughput. The plot of Fig.~\ref{fig7} reports the throughput of the GRAVITY instrument fed with the internal calibration source from January 2017 to July 2022. Having no particular information on the properties of the calibration lamp, we can at least observe two patterns with relatively stable throughput over the years. The jump towards a higher transmission in mid-2019 is unrelated to the IO chip, but rather corresponds to an increased throughput due to the change of grism. One can conclude that the system ``fibered integrating optics combiner'' does not appear to be the weak point of the whole transmission chain. The GRAVITY-like ABCD scheme due to the reduced number of pixels in comparison to the multi-axial scheme has motivated the MYSTIC team to consider an additional mode of their instrument using the spare version of the GRAVITY beam combiner. The tests are currently on-going at CHARA (J. Monnier, private communication). 
%----------------
%----------------
\\
%----------------
%----------------
%----------------
%----------------
\begin{figure}[h]
\centering
\includegraphics[width=0.60\textwidth]{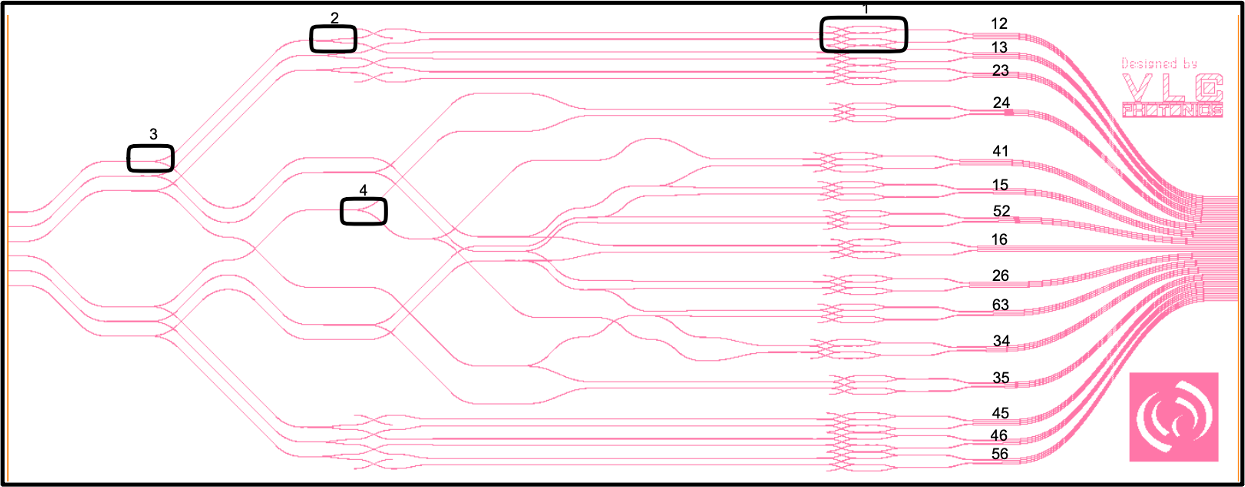}
\caption{Layout of the 6-telescope integrated beam combiner for a possible phase-delay fringe tracker at CHARA. A similar technology to the GRAVITY component is used. The chip contains 6 inputs and 60 encoded outputs and implements the ABCD scheme. Courtesy of D. Mourard.}\label{fig8}
\end{figure}\\
%----------------
%----------------
The success of the GRAVITY integrated optics beam combiner has motivated a first 6-telescope version with the a similar technology and manufactured by VLC-Photonics to serve as a phase-delay fringe tracker at CHARA to co-phase, for instance, the SPICA instrument. This solution, for which a first chip has been produced, is currently under evaluation (cf. Pannetier et al.\cite{Pannetier2020} and Mourard et al., this proceeding).

\subsubsection{Ultrafast-laser-written (ULI) beam combiner: extension to the K\,band}\label{io-beamcombiners}

The GRAVITY beam combiner -- and the PIONIER beam combiner in the H band before him -- is the prototype of science productive astrophotonics chip for interferometry. And probably the only one of this kind. In parallel, the simplicity and versatility of the ULI platform has motivated a team from the Heriot-Watt University, the University of Cologne and the AIP-Potsdam institutions to explore a new type of beam-combiner based on commercially available and highly transparent Infrasil-type subtrate of the silica family. The objective is to cover the whole K\,band with a glass substrate better matching the band of operation.\\
This instrumental research has led to the manufacturing of a two-telescope infrasil chip with photometric tapers fiber-in fiber-out. The accurate manufacturing optimization and lab characterization has produced a fairly achromatic chip delivering about 92\% of broadband interferomeric contrast in unpolarized light\cite{Benoit2021}\,.
%----------------
%----------------
\begin{figure}[b]
\centering
\includegraphics[width=0.80\textwidth]{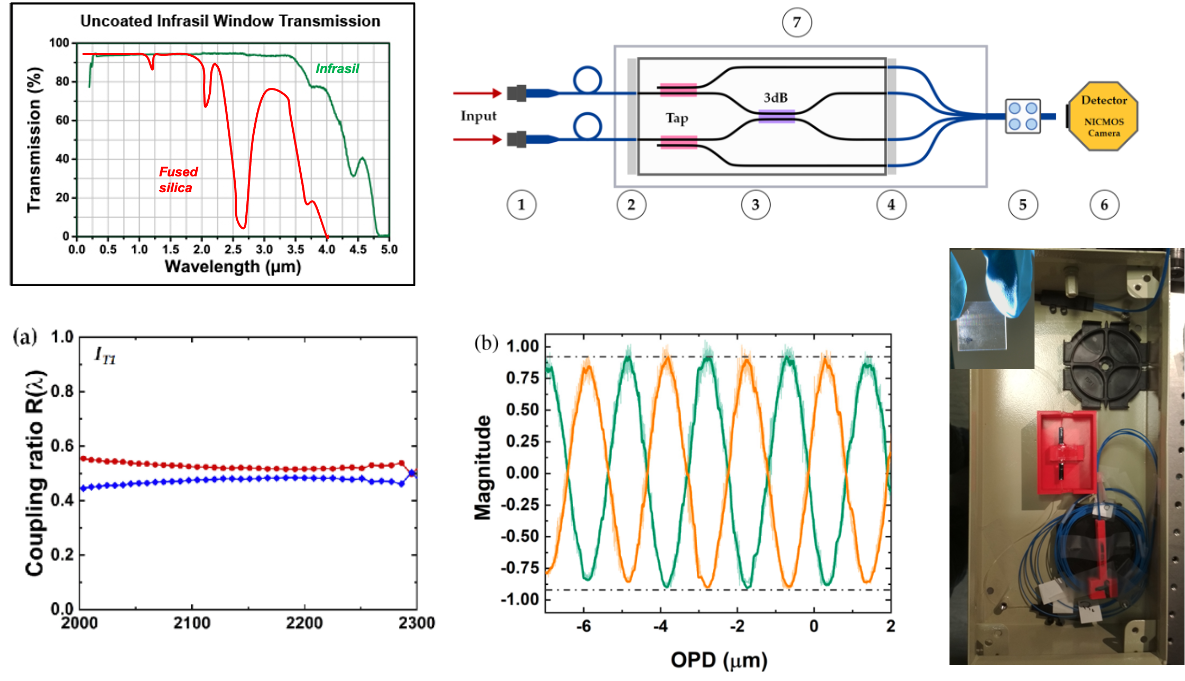}
\caption{Top/left: comparison of silica and infrasil transparency curves; Top/right: layout of the two-telescope combiner to be interfaced with the JouFlu infrastructure at CHARA; Bottom/left: achromatic splitting ratio across the K\,band; Bottom/center: zoom on the broadband interferometric contrast; Bottom/right: instrument packaging with a close view of the 11.7\,mm long chip in the inset.}\label{fig9}
\end{figure}
%----------------
%----------------
This first K\,band ULI combiner is being integrated and tested at CHARA with the JouFlu infrastructure and the NICMOS camera.

\subsubsection{Ultrafast-laser-written (ULI) beam combiner: the 4T-DBC experiment on-sky}

As an alternative scheme to the multi-axial and ABCD beam combination schemes, the Discrete Beam Combiner option was proposed\cite{Minardi2016} relying only on the evanescent coupling between channel waveguides, and thus avoiding all forms of bending losses in the interferometric combiner. This architecture relies on a 2D ``zig-zag'' configuration\cite{Diener2017} allowing non-nearest-neighbor interaction, which is essential to the encoding of the coherence function \cite{Minardi2015}\,, and which naturally requires the advantage of the 3D writing capabilities of ULI as opposed to planar integrated optics solutions. After several steps of laboratory characterization\cite{Nayak2020}\,, a four-telescope DBC was manufactured in borosilicate glass and tested at the William Herschel Telescope, in conjunction with a $\sim$30-40\% Strehl correction by the CANARY adaptive optics system serving as fringe-tracker. It is to be noted that an advantage of the DBC architecture is on the small numbers of encoded pixels in comparison to the ABCD or multi-axial scheme, in particular when increasing the number of apertures. For instance, for the multi-axial combination scheme, the minimum number of encoding pixels per wavelength channel is $\sim$30 in the four-telescope (4T) configuration and $\sim$140 in the 6T configuration, respectively. The pairwise ABCD combiner requires 24 pixels in the 4T and 60 pixels in the 6T configuration, while the DBC requires 23 pixels in the 4T and 41 pixels in the 6T configuration, respectively. Now the concept needed to be tested on sky.
\\
%----------------
%----------------
\begin{figure}[t]
\centering
\includegraphics[width=\textwidth]{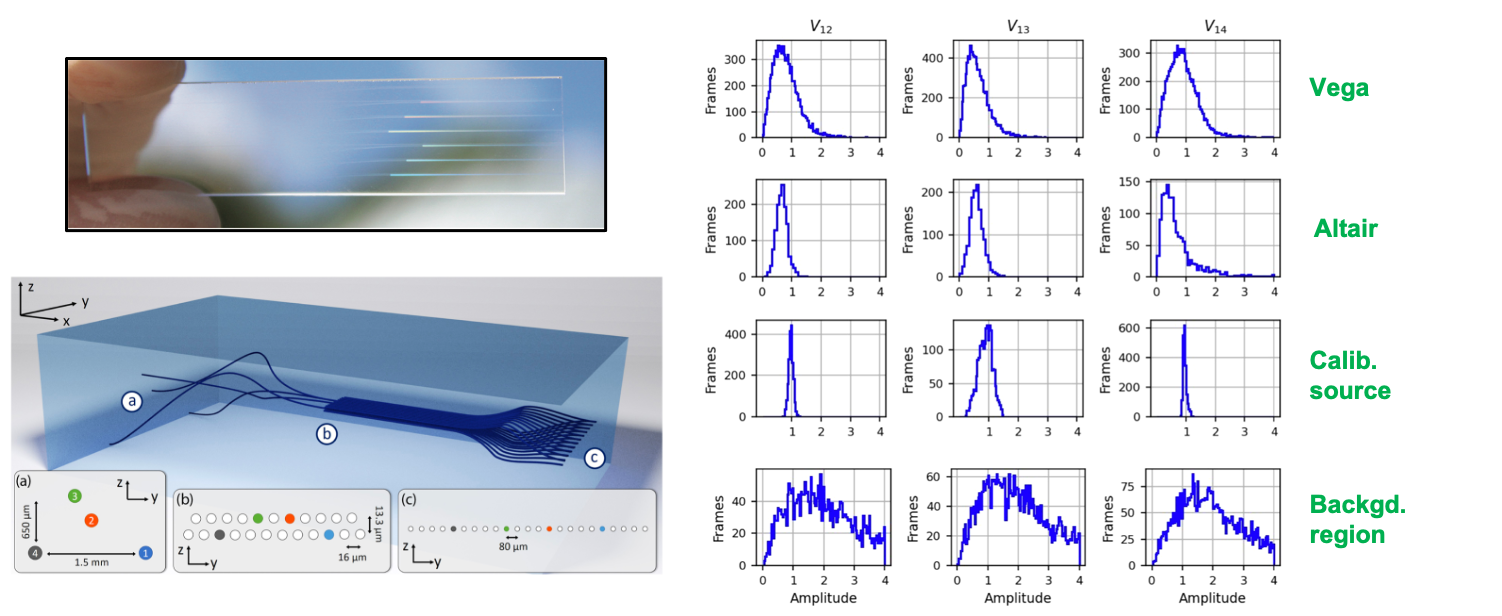}
\caption{Overview of the 4T-DBC experiment at the William Herschel Telescope. Left/bottom: design of the beam combiner with the remapping stage collecting flux from the four sub-apertures (1), the interferometric beam combiner (2), and the fan-out region to allow the read-out of the fluxes (3). Left/top: picture showing the typical microscope-blade compactness of the IO chip. Right: example of reconstructed visibilities when observing Vega, Altair and the internal calibration source. See Nayak et al.\cite{Nayak2021} for details.}\label{fig10}
\vspace{0.0cm}
\end{figure}\\
%----------------
%----------------
Light from Vega and Altair was coupled in the 4T-DBC combiner designed to fit in a classical aperture-masking/pupil remapping experiment (see Fig.~\ref{fig10}). The injection into the DBC component is controlled via a segmented mirror conjugated to a bulk-optics microlenses array. The retrieval of the coherence function was performed inverting the calibration V2PM matrix. Squared visibilities could be reconstructed for the two sources and compared to the internal calibration source, or to the retrieval from a noisy region on the detector. Instrumental visibilities of less than unity were retrieved on the six baselines, whereas the closure phases appeared paradoxically highly noisy for being self-calibrated quantities. It was concluded that, while the experiment could demonstrate the proof-of-principle of the on-sky DBC, the low flux conditions coupled to unavoidable long integration times had limited the yield of the experiment\cite{Nayak2021}\,. In a next step, the consideration of a 3D-printed lenslet array could significantly improve the stability of the transfer function.
  
\subsubsection{Ultrafast-laser-written (ULI) beam combiner: revival of nulling}

\begin{figure}[t]
\centering
\includegraphics[width=\textwidth]{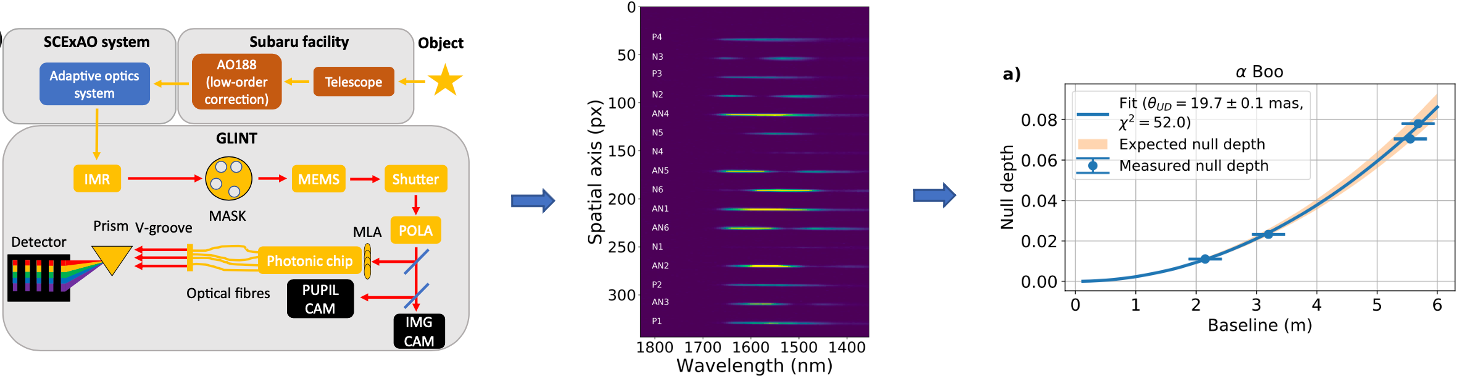}
\caption{Overview of the GLINT nuller operation in the H-band in dispersed mode\cite{Martinod2021}\,. In the center, the 12 nulled (N) and anti-nulled (AN) outputs are visible.}\label{fig11}
\vspace{0.0cm}
\end{figure}

In the last two years, nulling interferometry has experienced a new birth primarily thanks to the results of the GLINT instrument installed at Subaru (see Fig.~\ref{fig11}). GLINT relies on an extension of the original two-telescope prototype\cite{Norris2020b} to a four-input chip based on directional couplers and producing sixteen spectrally dispersed outputs corresponding to the six baselines (i.e., one null and one anti-null per coupler) and the four photometric channels. GLINT is interfaced through a remapping stage to the SCExAO output, while a MEMS (or segmented mirror) allowed to maximize the coupling and control the required phase shift to be on the dark fringe. In Martinod et al.\cite{Martinod2021}\,, the authors demonstrate in the lab an instrumental contrast of $\sim$10$^{-3}$ in dispersed light, while the on-sky observation of $\alpha$\,Boo in nulling mode has resulted in the detection of few 10$^{-2}$ stellar leakage, in agreement with the expected stellar diameter of this source. In a similar context, the potential of the ULI platform was further analyzed through simulations only by studying the potential of an integrated optics tri-coupler in an equilateral configuration within the interaction region to serve as achromatic nuller and in situ fringe-tracker simultaneously\cite{Martinod2021b}\,. These results open new perspectives for nulling interferometry on other observational platforms.

%----------------
%----------------
%----------------
%----------------

\subsubsection{The Kernel-Nulling approach: a lab demonstration}

\begin{figure}[t]
\centering
\includegraphics[width=0.9\textwidth]{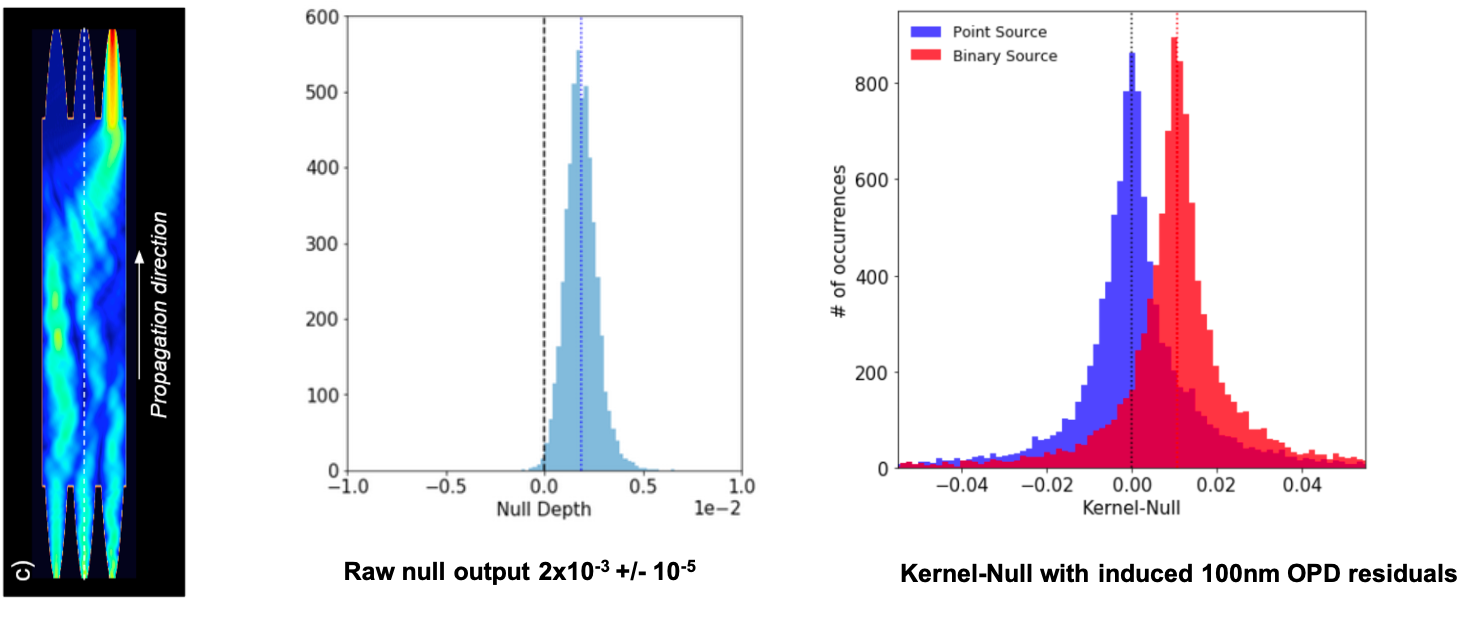}
\caption{Left: design of the 3-input MMI of the H band Kernel-nuller demonstrator; Center: distribution of the raw nulls; Right: effect of the self-calibrated kernel null for the detection of a faint companion and short separation. From Cvetojevic et al.\cite{Cvetojevic2022}\,.}\label{fig12}
\vspace{0.0cm}
\end{figure}

The Kernel-nulling approach based on self-calibrated quantities should allow the measurement of nulls that are less sensitive to phase excursions that are classically encountered in long-baseline interferometry from the ground\cite{Martinache2018}\,.
\\
An experimental validation of the technique has been demonstrated in the lab by Cvetojevic et al.\cite{Cvetojevic2022}\, making use of a three-input integrated optics multimode interferometer (MMI) design (see Fig.~\ref{fig12}). This is particularly relevant in the context of this astrophotonics review since the photonic chip is at the heart of this approach. 
In brief, three-input MMI produces one bright output, and two ``classical'' nulled outputs that are subtracted each other to form a kernel-null robust to random phase error from, for instance, the fringe-tracker (see their publication for details). The MMI photonic architecture was produced using UV-photolithography and operated essentially in the H-band. The main message is that, while raw nulls of 2$\times$10$^{-3}$ were measured, the kernel distribution resulting from the subtraction of the two nulled outputs proved to be a self-calibrated quantity with respect to an induced residual piston of 100\,nm rms. The kernel null was consistent with zero within an error of 10$^{-4}$, which has then permitted the detection of a simulated 10$^{-2}$ dimmer companion at a $\sim$2\,mas separation, assuming a VLTI configuration using the UTs an observing at the zenith. This laboratory demonstration marks an important step towards the assessment of nulling techniques from the ground. 

\newpage

%----------------
%----------------
%----------------
%----------------

\subsubsection{On-sky remapping instruments and on-chip active phase control}

The FIRST instrument dedicated to the remapping of the full telescope pupil in the visible and handling currently 2$\times$9 sub-apertures has now evolved towards a ``version 2'' in which a passive integrated optics chip achieves the beam combination instead of the original multi-axial encoding\cite{Barjot2020}\,. Furthermore, FIRST-v2 is the only hybrid case where it is attempted to insert a Niobate Lithium stage for active on-chip phase control\cite{Martin2020}\,, although the throughput seems still insufficient (E. Huby, provate communication).\\
Regarding this point, the availability of on-chip phase modulation holds a major interest in astrophotonics for the rapid stabilization of the phase in interferometry. While the Niobate Lithium technologies based on the electro-optic effect allow active phase control at the MHz speeds, other venues have been recently explored using the thermo-optic effect that operates at lower kHz speeds. In this area, I wish to briefly report on the promising result obtained by Montesinos-Ballester et al.\cite{Montesinos2019}\,, who demonstrated on-chip Fourier-Transform spectroscopy in the mid-infrared using an array of SiGe Mach-Zender Interferometers in which the OPD in one arm was scanned using the thermo-optic effect. This is certainly part of future improvements of all-photonic interferometric beam combiner chips. \\
Finally, a recent interesting concept that pertains to the area of astrophotonics interferometry is the ``waveguide-free'' approach proposed by Doelman et al.\cite{Doelman2021} to perform aperture-masking on large telescope and preventing baseline redundancy using holographic aperture masks made of liquid-crystal phase patterns. The idea is comparable to the segment-tilting experiment on the Keck by Monnier et al.\cite{Monnier2009}\,, but without having to unphase the primary mirror. All the possibly redundant baselines are encoded at different spatial locations on the detector taking advantage of the adequately dimensioned phase mask. Tested on-sky, the concept has remapped 83\% of the telescope pupil (30 segments out of 36), making it one of the most successful approach to efficient pupil remapping. 
%----------------
%----------------
\begin{figure}[h]
\centering
\includegraphics[width=\textwidth]{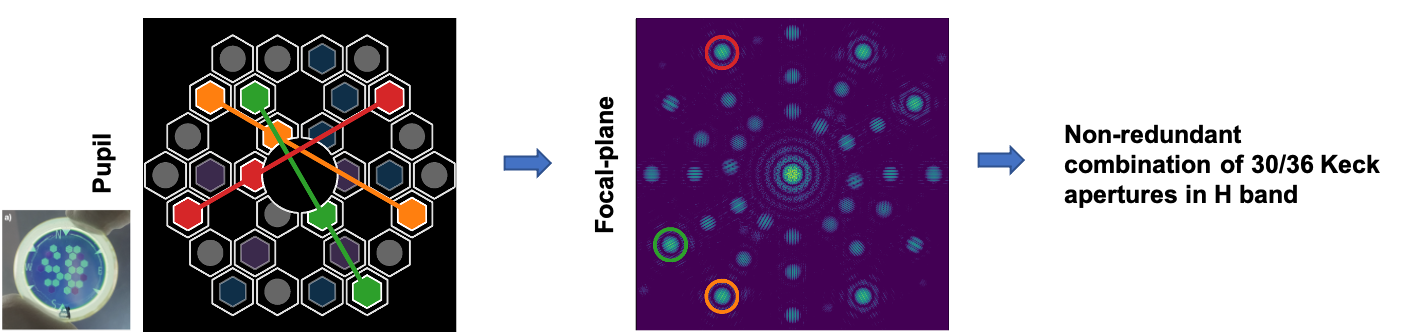}
\caption{Concept and physical holographic mask for the ``waveguide-free'' remapping of a telescope pupil\cite{Doelman2021}\,.}\label{fig13}
\vspace{0.0cm}
\end{figure}
%----------------
%----------------

\section{Discussion}\label{Discussion}

In this review, the recent progress in the field of astrophotonics for interferometry has been highlighted. No claim for full exhaustivity is made here, however the rapid growing of the field is emphasized. I did not treated the strong potential of astrophotonics for high-resolution spectroscopy and high-contrast imaging, but it is important to understand the strong complementarity and interplay between all these techniques in terms of photonic functionalities and fabrication platforms. These three observational techniques share common grounds as seen from the perspective of astrophotonics. 
I also did not emphasize in detail the recent progress in the field of mid-infrared photonics. However, I tried to cover these themes in a more general vision that is summarized in Fig.~\ref{fig3}. The future of up-conversion technologies pioneered by the group of Fran\c cois Reynaud (see for instance Lehmann et al.\cite{Lehmann2019b}) may also bring further novelty, unfortunately not addressed here. Finally, I fully concentrated here on prospects for ground-based observations, whereas it is likely that one the strongest potential of astrophotonics is in space due to obvious small-scale integration capabilities. Groups outside the immediate astronomical environment have started to explore the impact of high-energy radiation on waveguide structures, as this might be expected from a low Earth orbit space environment\cite{Piacentini2021}\,.\\
\\
Interferometry and spectroscopy are currently two important drivers for astrophotonics and, given the conference, emphasis has been set on the prospects for the ground-based interferometers like the VLTI or CHARA. However, it is easy to predict that a new frontier for astrophotonics will be set by the emergence of the class of Extremely Large Telescope for which new innovation will be needed: considering that the primary mirror of the E-ELT is formed by 900 segments each with a diameter of 1.4\,m, the task for interferometry-biased astrophotonicists may result arduous, requiring significant creativity.\\
\\
A last comment regards the wider recognition of the field. The recent Decadal Survey ``Pathways to Discovery in Astronomy and Astrophysics for the 2020s'' has evoked for the first time the discipline of astrophotonics, highlighting the importance of a measured plan. The report quotes in the section K.4.7 Technology development -- Astrophotonics: {\it  
Strengthening the coordination between the most active astrophotonics research groups in the United States would optimize resources and facilitate the passage from laboratory research to industrial partnership. This could be done through the creation of a distributed, multi-disciplinary Institute of Astrophotonics to coordinate the teams working in this field. The more coordinated approach adopted by Europe (Germany in particular) and Australia has led to success and leadership in this field. A few tens- of-millions of dollars of funding over the next decade would be needed to significantly advance this technology and reestablish U.S. leadership in astrophotonics. 
}\\
The need for an Institute of Astrophotonics might be questionable, but the community is definitely present and active.

% Glued microlenses 

%\section{Conclusions}

%%%%%%%%%%%%%%%%%%%%%%%%%%%%%%%%%%%%%%%%%%%%%%%%%%%%%%%%%%%%%

%%%%%%%%%%%%%%%%%%%%%%%%%%%%%%%%%%%%%%%%%%%%%%%%%%%%%%%%%%%%%
\acknowledgments          
L. Labadie acknowledges discussions with many colleagues in the field and with interested persons and students. 
 %%%%%%%%%%%%%%%%%%%%%%%%%%%%%%%%%%%%%%%%%%%%%%%%%%%%%%%%%%%%%
%%%%% References %%%%%

\bibliography{report}   %>>>> bibliography data in report.bib
\bibliographystyle{spiebib}   %>>>> makes bibtex use spiebib.bst

\end{document}